\documentclass{article}

\usepackage{color}
\usepackage{amsmath, amssymb}

\newcommand{\RM}{\mathbb{R}}
\newcommand{\ZM}{\mathbb{Z}}

\newcommand{\CM}{\mathbb{C}}
\newtheorem{theorem}{Theorem}

\newtheorem{cor}{Corollary}

\begin{document}

\title{{\bf An analogue of the Riemann Hypothesis \\
via quantum walks}
\vspace{15mm}}

\author{Norio KONNO \\
Department of Applied Mathematics, Faculty of Engineering \\ 
Yokohama National University \\
Hodogaya, Yokohama, 240-8501, Japan \\
e-mail: konno-norio-bt@ynu.ac.jp \\
}

\date{\empty }

\maketitle

\vspace{50mm}


\vspace{20mm}











\clearpage

\begin{abstract}
We consider an analogue of the well-known Riemann Hypothesis based on quantum walks on graphs with the help of the Konno-Sato theorem. Furthermore, we give some examples for complete, cycle, and star graphs.
\end{abstract}

\vspace{10mm}

\begin{small}
\par\noindent
{\bf Keywords}: Zeta function, Riemann Hypothesis, Quantum walk, Grover walk, Konno-Sato theorem
\end{small}

\vspace{10mm}

\section{Introduction \label{sec01}}
Inspired our recent work for a series of Zeta/Correspondence \cite{K0, K1, K2, K3, K4, K5, K7, K8, K6} on the relation between zeta functions and some models such as random walk (RW) and quantum walk (QW), we consider an analogue of the well-known Riemann Hypothesis via QWs on graphs with the help of the Konno-Sato theorem. The QW is a quantum counterpart of the RW. The Konno-Sato theorem \cite{KonnoSato} treats the Grover walk which is one of the most well-investigated QWs. Concerning QW, see \cite{Konno2008, ManouchehriWang, Portugal, Venegas}, as for RW, see \cite{Norris, Spitzer}, and as for the Riemann Hypothesis, see \cite{Edwards, GM, Terras}, for examples. More precisely, we introduce a new zeta function $\Lambda^{QW} _G (s) = \det \left( {\bf M} - s (1-s) {\bf I} \right)$ for a suitable matrix ${\bf M}$ on a simple connected graph $G$ by using the Konno-Sato theorem, where $s$ is a complex number and ${\bf I}$ is the identity matrix. Here $\Lambda^{QW} _G (s)$ corresponds the completed zeta function  $\Lambda (s)$. As for the detailed definition, see Section \ref{sec03}. Moreover, we prove that $\Lambda^{QW} _G (s)$ satisfies the Functional Equation: $\Lambda^{QW} _G (s) = \Lambda^{QW} _G (1- s)$ and the Riemann Hypothesis: if $\rho \in {\rm Zero} (\Lambda^{QW} _G)$, then $\Re (\rho) = 1/2$, where ${\rm Zero} (f)$ is the set of the zeros of $f(s)$ and $\Re (z)$ is the real part of $z$. Remark that the original Riemann Hypothesis: if $\rho \in {\rm Zero} (\Lambda)$, then $\Re (\rho) = 1/2$ for the completed zeta function  $\Lambda (s)$. Equivalently, if $\rho \in {\rm Zero} (\zeta)$ with $0 < \Re (\rho) <1$, then $\Re (\rho) =1/2$ for Riemann's zeta function $\zeta (s) = \sum_{n=1}^{\infty} 1/n^s$. The present manuscript is the first step of the study on a connection between the Riemann Hypothesis and the QW.

The rest of this paper is organized as follows. Section \ref{sec02} gives a brief overview of the Konno-Sato theorem. In Section \ref{sec03}, we explain an analogue of the Riemann Hypothesis in our setting. Section \ref{sec04} presents some examples for complete, cycle, and star graphs. Finally, Section \ref{sec05} is devoted to conclusion.

\section{Konno-Sato Theorem \label{sec02}}
First we introduce the following notation: $\mathbb{Z}$ is the set of integers, $\mathbb{Z}_{>} = \{1,2,3, \ldots \}$,  $\mathbb{R}$ is the set of real numbers, and $\mathbb{C}$ is the set of complex numbers.

In this section, we briefly review the Konno-Sato theorem given by \cite{KonnoSato}. This theorem treats a relation for eigenvalues between QWs and RWs. More specifically, the Grover walk (which is QW determined by the Grover matrix) with flip-flop shift type (called F-type) and simple symmetric RW (whose walker jumps to each of its nearest neighbors with equal probability) on a graph. We assume that all graphs are simple.

Let $G=(V(G),E(G))$ be a connected graph (without multiple edges and loops) with the set $V(G)$ of vertices and the set $E(G)$ of unoriented edges $uv$ joining two vertices $u$ and $v$. Moreover, let $n=|V(G)|$ and $m=|E(G)|$ be the number of vertices and edges of $G$, respectively. For $uv \in E(G)$, an arc $(u,v)$ is the oriented edge from $u$ to $v$. Let $D(G)$ be the symmetric digraph corresponding to $G$, i.e., $D(G)= \{ (u,v),(v,u) \mid uv \in E(G) \}$. For $e=(u,v) \in D(G)$, set $u=o(e)$ and $v=t(e)$. Furthermore, let $e^{-1}=(v,u)$ be the {\em inverse} of $e=(u,v)$. For $v \in V(G)$, the {\em degree} $\deg {}_G \ v = \deg v = d_v $ of $v$ is the number of vertices adjacent to $v$ in $G$. If $ \deg {}_G \ v=k$ (constant) for each $v \in V(G)$, then $G$ is called {\em $k$-regular}. A {\em path $P$ of length $n$} in $G$ is a sequence $P=(e_1, \ldots ,e_n )$ of $n$ arcs such that $e_i \in D(G)$, $t( e_i )=o( e_{i+1} ) \ (1 \leq i \leq n-1)$. If $e_i =( v_{i-1} , v_i )$ for $i=1, \cdots , n$, then we write $P=(v_0, v_1, \ldots ,v_{n-1}, v_n )$. Put $ \mid P \mid =n$, $o(P)=o( e_1 )$ and $t(P)=t( e_n )$. Also, $P$ is called an {\em $(o(P),t(P))$-path}. We say that a path $P=( e_1 , \ldots , e_n )$ has a {\em backtracking} if $ e^{-1}_{i+1} =e_i $ for some $i \ (1 \leq i \leq n-1)$. A $(v, w)$-path is called a {\em $v$-cycle} (or {\em $v$-closed path}) if $v=w$. Let $B^r$ be the cycle obtained by going $r$ times around a cycle $B$. Such a cycle is called a {\em multiple} of $B$. A cycle $C$ is {\em reduced} if both $C$ and $C^2 $ have no backtracking. The {\em Ihara zeta function} of a graph $G$ is a function of a complex variable $u$ with $|u|$ sufficiently small, defined by 
\begin{align*}
{\bf Z} (G, u)= \exp \left( \sum^{\infty}_{r=1} \frac{N_r}{r} u^r \right), 
\end{align*}
where $N_r$ is the number of reduced cycles of length $r$ in $G$. Let $G$ be a simple connected graph with $n$ vertices $v_1, \ldots ,v_n $. The {\em adjacency matrix} ${\bf A}_n = [a_{ij} ]$ is the $n \times n$ matrix such that $a_{ij} =1$ if $v_i$ and $v_j$ are adjacent, and $a_{ij} =0$ otherwise. The following result was obtained by Ihara \cite{Ihara} and Bass \cite{Bass}.
\begin{theorem}[Ihara \cite{Ihara}, Bass \cite{Bass}]
Let $G$ be a simple connected graph with $V(G)= \{ v_1 , \ldots , v_n \}$ and $m$ edges. Then we have
\begin{align*}
{\bf Z} (G,u )^{-1} =(1- u^2 )^{\gamma-1} 
\det \left( {\bf I}_n -u {\bf A}_n + u^2 ( {\bf D}_n - {\bf I}_n ) \right). 
\end{align*}
Here $\gamma$ is the Betti number of $G$ {\rm (}i.e., $\gamma = m - n +1${\rm )}, ${\bf I}_n$ is the $n \times n$ identity matrix, and ${\bf D}_n = [d_{ij}]$ is the $n \times n$ diagonal matrix with $d_{ii} = \deg v_i$ and $d_{ij} =0 \ (i \neq j)$. 
\end{theorem}

Let $G$ be a simple connected graph with $V(G)= \{ v_1 , \ldots , v_n \}$ and $m$ edges. Set $d_j = d_{v_j} = \deg v_j \ (j=1, \ldots , n)$. Then the $2m \times 2m$ {\em Grover matrix} ${\bf U}_{2m} = [ U_{ef} ]_{e,f \in D(G)} $ of $G$ is defined by 
\begin{align*}
U_{ef} =\left\{
\begin{array}{ll}
2/d_{t(f)} (=2/d_{o(e)} ) & \mbox{if $t(f)=o(e)$ and $f \neq e^{-1} $, } \\
2/d_{t(f)} -1 & \mbox{if $f= e^{-1} $, } \\
0 & \mbox{otherwise. }
\end{array}
\right. 
\end{align*}
The discrete-time QW with the Grover matrix ${\bf U}_{2m}$ as a time evolution matrix is the Grover walk with F-type on $G$. Then the $n \times n$ matrix ${\bf P}_{n} = [ P_{uv} ]_{u,v \in V(G)}$ is given by
\begin{align*}
P_{uv} =\left\{
\begin{array}{ll}
1/( \deg {}_G \ u)  & \mbox{if $(u,v) \in D(G)$, } \\
0 & \mbox{otherwise.}
\end{array}
\right.
\end{align*}
Note that the matrix ${\bf P}_n$ is the transition probability matrix of the simple symmetric RW on $G$. We introduce the {\em positive support} ${\bf F}^+ = [ F^+_{ij} ]$ of a real matrix ${\bf F} = [ F_{ij} ]$ as follows: 
\begin{align*}
F^+_{ij} =\left\{
\begin{array}{ll}
1 & \mbox{if $F_{ij} >0$, } \\
0 & \mbox{otherwise}.
\end{array}
\right.
\end{align*}
Ren et al. \cite{RenEtAl} showed that the Perron-Frobenius operator (or edge matrix) of a graph is the positive support $({}^{\rm{T}}{\bf U}_{2m})^+ $ of the transpose of its Grover matrix ${\bf U}_{2m}$, i.e., 
\begin{align*}
{\bf Z} (G,u)^{-1} = \det \left( {\bf I}_{2m} -u( {}^{\rm{T}}{\bf U}_{2m})^+ \right)= \det \left( {\bf I}_{2m} -u {\bf U}_{2m} ^+ \right). 
\end{align*}
The Ihara zeta function of a graph $G$ is just a zeta function on the positive support of the Grover matrix of $G$. That is, the Ihara zeta function corresponds to the positive support version of the Grover walk (defined by the positive support of the Grover matrix ${\bf U}_{2m} ^+$) with F-type on $G$.

Now we propose another zeta function of a graph. Let $G$ be a simple connected graph with $m$ edges. Then we define a zeta function $ \overline{{\bf Z}} (G, u)$ of $G$ satisfying 
\begin{align*}
\overline{{\bf Z}} (G, u)^{-1} = \det ( {\bf I}_{2m} -u {\bf U}_{2m} ).    
\end{align*}
In other words, this zeta function corresponds to the Grover walk (defined by the Grover matrix ${\bf U}_{2m}$) with F-type on $G$.

In this setting, Konno and Sato \cite{KonnoSato} presented the following result which is called the {\em Konno-Sato theorem}. 

\begin{theorem}[Konno and Sato \cite{KonnoSato}]
Let $G$ be a simple connected graph with $n$ vertices and $m$ edges. Then  
\begin{align}  
\overline{{\bf Z}} (G, u)^{-1} 
= \det ( {\bf I}_{2m} - u {\bf U}_{2m} )
=(1-u^2)^{m-n} \det \left( (1+u^2) {\bf I}_{n} -2u {\bf P}_n \right).
\label{wakatakakage1a}
\end{align}
\label{KS} 
\end{theorem}
If we take $u = 1/\lambda$, then Eq. \eqref{wakatakakage1a} implies
\begin{align}
\det \left( \lambda {\bf I}_{2m} - {\bf U}_{2m} \right)
= ( \lambda {}^2 -1)^{m-n} \det \left( ( \lambda {}^2 +1) {\bf I}_n -2 \lambda {\bf P}_n \right).
\label{wakatakakage00}
\end{align}
Furthermore, Eq. \eqref{wakatakakage00} can be rewritten as 
\begin{align}
\det \left( \lambda {\bf I}_{2m} - {\bf U}_{2m} \right)= ( \lambda {}^2 -1)^{m-n} \times \prod_{ \lambda {}_{{\bf P}_n}  \in {\rm Spec} ({\bf P}_n)} \left( \lambda {}^2 +1 -2 \lambda {}_{{\bf P}_n} \lambda \right), 
\label{star01}
\end{align}
where ${\rm Spec} ({\bf B})$ is the set of eigenvalues of a square matrix ${\bf B}$. More precisely, we also use the following notation: 
\begin{align*}
{\rm Spec} ({\bf B}) = \left\{ \left[ \lambda_1 \right]^{l_1}, \ \left[ \lambda_2 \right]^{l_2}, \ \ldots \ , \left[ \lambda_k \right]^{l_k} \right\},
\end{align*}
where $\lambda_j$ is the eigenvalue of ${\bf B}$ and $l_j \in \ZM_{>}$ is the multiplicity of $\lambda_j$ for $j=1,2, \ldots, k$. Set $|{\rm Spec} ({\bf B}) | = l_1 + l_2 + \cdots + l_k$. It follows from $\lambda {}^2 +1 -2 \lambda {}_{{\bf P}_n} \lambda =0$ that $\lambda{}_{{\bf U}_{2m}} \in {\rm Spec} ({\bf U}_{2m})$ is given by
\begin{align}
\lambda{}_{{\bf U}_{2m}} = \lambda {}_{{\bf P}_n} \pm i \sqrt{1- \lambda {}^2_{{\bf P}_n}}.
\label{star02}
\end{align}
Remark that $\lambda {}_{{\bf P}_n} \in [-1,1]$. Noting Eqs. \eqref{star01} and \eqref{star02}, we introduce ${\rm Spec} \left( {\bf U}_{2m} : {\rm RW} \right)$ and ${\rm Spec} \left( {\bf U}_{2m} : {\rm RW}^c \right)$ as follows:  
\begin{align*}
&{\rm Spec} \left( {\bf U}_{2m} : {\rm RW} \right) 
\\
& \qquad = \left\{ \left[ \lambda {}_{{\bf P}_n} + i \sqrt{1- \lambda {}^2_{{\bf P}_n}} \right]^1,  \ \left[ \lambda {}_{{\bf P}_n} - i \sqrt{1- \lambda {}^2_{{\bf P}_n}} \right]^1  \ : \  \lambda {}_{{\bf P}_n} \in {\rm Spec} ({\bf P}_n) \right\},
\\
&{\rm Spec} \left( {\bf U}_{2m} : {\rm RW}^c \right) 
= \left\{ \left[ 1 \right]^{|m-n|},  \ \left[ -1 \right]^{|m-n|} \right\}.
\end{align*}
When $m=n$, we let ${\rm Spec} \left( {\bf U}_{2m} : {\rm RW}^c \right) = \emptyset$. Note that $|{\rm Spec} \left( {\bf U}_{2m} : {\rm RW} \right)| = 2n$ and $|{\rm Spec} \left( {\bf U}_{2m} : {\rm RW}^c \right)| = 2 |m-n|$. We should remark that ${\rm Spec} \left( {\bf U}_{2m} : {\rm RW} \right)$ corresponds to the eigenvalue of ${\bf P}_n$ which is the transition probability matrix of the simple symmetric RW on $G$. On the other hand, ${\rm Spec} \left( {\bf U}_{2m} : {\rm RW}^c \right)$ does not corresponds to the RW, so superscript ``c" (of ${\rm RW}^c$) stands for ``complement". Therefore we obtain

\begin{cor}
Let $G$ be a simple connected graph with $n$ vertices and $m$ edges. 
\par
\
\par\noindent
{\rm (i)} If $m > n$, then   
\begin{align*}
{\rm Spec} ({\bf U}_{2m}) = {\rm Spec} \left( {\bf U}_{2m} : {\rm RW} \right) \cup {\rm Spec} \left( {\bf U}_{2m} : {\rm RW}^c \right).
\end{align*}
\par\noindent
{\rm (ii)} If $m = n$, then   
\begin{align*}
{\rm Spec} ({\bf U}_{2m}) = {\rm Spec} \left( {\bf U}_{2m} : {\rm RW} \right).
\end{align*}
\par\noindent
{\rm (iii)} If $m < n$, then   
\begin{align*}
{\rm Spec} ({\bf U}_{2m}) = {\rm Spec} \left( {\bf U}_{2m} : {\rm RW} \right) \setminus {\rm Spec} \left( {\bf U}_{2m} : {\rm RW}^c \right).
\end{align*}
\label{moderuna011}
\end{cor}
\par
\
\par
In Section \ref{sec04}, we will give some examples for each case. Since ${\bf U}_{2m}$ is unitary, $\lambda \in {\rm Spec} ({\bf U}_{2m})$ satisfies $|\lambda|=1$, so we put $\lambda = e^{i \theta} \ (\theta \in [0, 2 \pi))$. Thus we get
\begin{align}
\cos \theta = \frac{\lambda + \overline{\lambda}}{2},
\label{wakatakakage01}
\end{align}
where $\overline{\lambda}$ is the complex conjugate of $\lambda \in \CM.$ This is called the Joukowsky transform. It follows from  Eq. \eqref{wakatakakage01} and $\lambda^{-1} = \overline{\lambda}$ that Eq. \eqref{wakatakakage00} becomes 
\begin{align*}
\det \left( \lambda {\bf I}_{2m} - {\bf U}_{2m} \right)
= ( \lambda {}^2 -1 )^{m-n} ( 2 \lambda)^{n} \det \left( \cos \theta \cdot {\bf I}_n - {\bf P}_n \right).
\end{align*}
Therefore we have a relation ``$\cos \theta \in {\rm Spec} ({\bf P}_n)$'' \ $\Longrightarrow$ \ ``$\lambda = e^{i \theta} \in {\rm Spec} ({\bf U}_{2m})$'' which is sometimes called the {\em spectral mapping theorem} in the study of QW (see \cite{SS}, for example). We will explain in a more detailed fashion. We set
\begin{align*}
{\rm Spec} ({\bf P}_n) = \left\{ \left[ \cos \theta_1 \right]^{l_1}, \ \left[ \cos \theta_2 \right]^{l_2}, \ \ldots \ , \left[ \cos \theta_p \right]^{l_p} \right\},
\end{align*}
where $|{\rm Spec} ({\bf P}_n) | = l_1 + l_2 + \cdots + l_p = n$ and $0 = \theta_1 < \theta_2 < \cdots < \theta_p < 2 \pi$. Then we get
\begin{align}
&{\rm Spec} \left( {\bf U}_{2m} : {\rm RW} \right) 
\nonumber
\\
& \qquad = \left\{ \left[ e^{i \theta_1} \right]^{l_1}, \ \left[ e^{- i \theta_1} \right]^{l_1}, \ \left[ e^{i \theta_2} \right]^{l_2}, \ \left[ e^{- i \theta_2} \right]^{l_2}, \ \ldots \ , \left[ e^{i \theta_p} \right]^{l_p}, \ \left[ e^{- i \theta_p} \right]^{l_p} \right\}.
\label{star10}
\end{align}
Remark that $|{\rm Spec} \left( {\bf U}_{2m} : {\rm RW} \right) | = 2 \left( l_1 + l_2 + \cdots + l_p \right) = 2n$.

\section{Analogue of the Riemann Hypothesis \label{sec03}}
The {\em Riemann zeta function} for $s \in \CM$ with $\Re (s)>1$ is defined by
\begin{align*}
\zeta (s) = \sum_{n=1}^{\infty} \frac{1}{n^s} = \prod_{p :{\rm prime}} \left( 1 - p^{-s} \right)^{-1},
\end{align*}
where $\Re (z)$ is the real part of $z \in \CM$. It is known that $\zeta (s)$ has a meromorphic continuation to the entire complex plane. Moreover the {\em completed zeta function} $\Lambda (s)$ is given by
\begin{align*}
\Lambda (s) = \pi^{-s/2} \ \Gamma \left( \frac{s}{2} \right) \zeta (s),
\end{align*}
where $\Gamma (z)$ is the gamma function (see \cite{Andrews1999}, for example). Then $\Lambda (s)$ satisfies the {\bf Functional Equation}:  
\begin{align*}
\Lambda (s) = \Lambda (1-s).
\end{align*}
The {\bf Riemann Hypothesis}, which is stated in terms of $\Lambda$, is the following:
\begin{align*}
\Re (\rho) = \frac{1}{2} \qquad (\rho \in {\rm Zero} (\Lambda)),
\end{align*}
where 
\begin{align*}
{\rm Zero} (\Lambda) = \left\{ s \in \CM : \Lambda (s) = 0 \right\}.
\end{align*}
Concerning the Riemann Hypothesis, see \cite{Edwards, GM, Terras}, for instance. In this background, we want to find the following zeta function $\Lambda^{QW} _G (s)$ via the QW for a graph $G$ as a counterpart of completed zeta function $\Lambda (s)$ such as 
\par
\
\par\noindent
(i) {\bf Functional Equation}:
\begin{align*}
\Lambda^{QW} _G (s) = \Lambda^{QW} _G (1- s).  
\end{align*}
\par\noindent
(ii) {\bf Riemann Hypothesis}:
\begin{align*}
\Re (\rho) = \frac{1}{2} \qquad (\rho \in {\rm Zero} (\Lambda^{QW} _G)),
\end{align*}
where 
\begin{align*}
{\rm Zero} (\Lambda^{QW} _G) = \left\{ s \in \CM : \Lambda^{QW} _G (s) = 0 \right\}.
\end{align*}
That is, 
\begin{align*}
\text{if} \quad \Lambda^{QW} _G (\rho) = 0, \quad \text{then} \quad \Re (\rho) = \frac{1}{2}.
\end{align*}
To do so, we use the Konno-Sato theorem as described below. We assume that $G$ is a simple connected graph with $n$ vertices and $m$ edges. Put $\lambda \in {\rm Spec} ({\bf U}_{2m})$, where $\lambda = e^{i \theta}$ with $\theta \in [0, 2 \pi)$. In the complex plane, $\lambda$ corresponds to a point $(\cos \theta, \sin \theta)$. Let $\ell$ be the line through the points $(\cos \theta, \sin \theta)$ and $(1,0)$. If we consider the complex plane as $xy$ plane, then the equation for line $\ell$ becomes 
\begin{align}
y = - \frac{\sin \theta}{1 - \cos \theta} \ (x-1) \qquad (\theta \in (0, 2 \pi)).
\label{wakatakakage12}
\end{align}
When $x=1/2$, Eq. \eqref{wakatakakage12} yields the value
\begin{align*}
y = 
\frac{1}{2} \cot \left( \frac{\theta}{2} \right), 
\end{align*}
for $\theta \in (0, 2 \pi)$, so the point $\left( 1/2, \cot(\theta/2)/2 \right)$ lies on $\ell$. Thus, $\cot(\theta/2)/2$ is the $y$ coordinate of the intersection of $\ell$ and $x=1/2$. Therefore, we want to relate $\lambda = e^{i \theta} \in {\rm Spec} ({\bf U}_{2m})$ with $\theta \in (0, 2 \pi)$ to $\rho = \rho (\theta) \in {\rm Zero} (\Lambda^{QW} _G)$ in the following way:
\begin{align}
\rho = \rho (\theta) = \frac{1}{2} + \frac{i}{2} \cot \left( \frac{\theta}{2} \right) \qquad (\theta \in (0, 2 \pi)).
\label{wakatakakage02}
\end{align}
On the other hand, if $\theta = 0,$ then we put 
\begin{align}
\rho = \rho (0) = \frac{1}{2} + i \cdot (+ \infty).
\label{wakatakakage02b}
\end{align}
Noting Eqs. \eqref{star10}, \eqref{wakatakakage02} and \eqref{wakatakakage02b}, we introduce the following sets ${\rm Zero} \left( \Lambda_G^{QW} : {\rm RW} \right)$ and ${\rm Zero} \left( \Lambda_G^{QW} : {\rm RW}^c \right)$ corresponding to ${\rm Spec} \left( {\bf U}_{2m} : {\rm RW} \right)$ and ${\rm Spec} \left( {\bf U}_{2m} : {\rm RW}^c \right)$, respectively.
\begin{align*}
&{\rm Zero} \left( \Lambda_G^{QW} : {\rm RW} \right) 
\\
& \qquad = \left\{ \left[ \rho (\theta_1) \right]^{l_1}, \ \left[ \rho (- \theta_1) \right]^{l_1}, \ \left[ \rho (\theta_2) \right]^{l_2}, \ \left[ \rho (- \theta_2) \right]^{l_2}, \ \ldots \ , \left[ \rho (\theta_p) \right]^{l_p}, \ \left[ \rho (- \theta_p) \right]^{l_p} \right\}
\\
& \qquad = \left\{ \ \left[ \frac{1}{2} + \frac{i}{2} \cot \left( \frac{\theta_q}{2} \right) \right]^{l_q}, \ \left[ \frac{1}{2} - \frac{i}{2} \cot \left( \frac{\theta_q}{2} \right) \right]^{l_q} \ : \ q=1,2, \ldots,p \ \right\},
\\
&{\rm Zero} \left( \Lambda_G^{QW} : {\rm RW}^c \right)
\\  
& \qquad = \left\{ \left[ \rho (0) \right]^{|m-n|}, \ \left[ \rho (\pi) \right]^{|m-n|} \right\} 
= \left\{ \ \left[ \frac{1}{2} + i \cdot ( + \infty) \right]^{|m-n|}, \ \left[ \frac{1}{2} \right]^{|m-n|} \right\}.
\end{align*}
Remark that $| {\rm Zero} \left( \Lambda_G^{QW} : {\rm RW} \right) | = 2 \left( l_1 + l_2 + \cdots + l_p \right) = 2n$ and $| {\rm Zero} \left( \Lambda_G^{QW} : {\rm RW}^c \right) | = 2 |m-n|$. Furthermore, we introduce ${\rm Zero} \left( \Lambda_G^{QW} \right)$ as follows: 
\par\noindent
{\rm (i)} If $m > n$, then   
\begin{align*}
{\rm Zero} \left( \Lambda_G^{QW} \right) 
&= {\rm Zero} \left( \Lambda_G^{QW} : {\rm RW} \right) \cup {\rm Zero} \left( \Lambda_G^{QW} : {\rm RW}^c \right)
\\
&= \left\{ \ \left[ \frac{1}{2} + \frac{i}{2} \cot \left( \frac{\theta_q}{2} \right) \right]^{l_q}, \ \left[ \frac{1}{2} - \frac{i}{2} \cot \left( \frac{\theta_q}{2} \right) \right]^{l_q} \ : \ q=1,2, \ldots, p \ \right\} 
\\
&\cup \left\{ \ \left[ \frac{1}{2} + i \cdot ( + \infty) \right]^{m-n}, \ \left[ \frac{1}{2} \right]^{m-n} \right\}.
\end{align*}
\par\noindent
{\rm (ii)} If $m = n$, then   
\begin{align*}
{\rm Zero} \left( \Lambda_G^{QW} \right) 
&= {\rm Zero} \left( \Lambda_G^{QW} : {\rm RW} \right) 
\\
&= \left\{ \ \left[ \frac{1}{2} + \frac{i}{2} \cot \left( \frac{\theta_q}{2} \right) \right]^{l_q}, \ \left[ \frac{1}{2} - \frac{i}{2} \cot \left( \frac{\theta_q}{2} \right) \right]^{l_q} \ : \ q=1,2, \ldots, p \ \right\}. 
\end{align*}
\par\noindent
{\rm (iii)} If $m < n$, then   
\begin{align*}
{\rm Zero} \left( \Lambda_G^{QW} \right) 
&= {\rm Zero} \left( \Lambda_G^{QW} : {\rm RW} \right) \setminus {\rm Zero} \left( \Lambda_G^{QW} : {\rm RW}^c \right)
\\
&= \left\{ \ \left[ \frac{1}{2} + \frac{i}{2} \cot \left( \frac{\theta_q}{2} \right) \right]^{l_q}, \ \left[ \frac{1}{2} - \frac{i}{2} \cot \left( \frac{\theta_q}{2} \right) \right]^{l_q} \ : \ q=1,2, \ldots, p \ \right\} 
\\
&\setminus \left\{ \ \left[ \frac{1}{2} + i \cdot ( + \infty) \right]^{n-m}, \ \left[ \frac{1}{2} \right]^{n-m} \right\}.
\end{align*}

Noting Eq. \eqref{wakatakakage02}, we define $\Lambda_G^{QW} (s)$ by 
\begin{align}
\Lambda_G^{QW} (s) = \det \left( {\bf M}_n - s (1-s) {\bf I}_{n} \right),
\label{takayasu01}
\end{align}
for a suitable $n \times n$ matrix ${\bf M}_n$. By Eq. \eqref{takayasu01}, we easily confirm that $\Lambda_G^{QW} (s)$ satisfies $\Lambda_G^{QW} (s) = \Lambda_G^{QW} (1-s)$, i.e., Functional Equation. On the other hand, $\Lambda_G^{QW} (s)$ can be rewritten as 
\begin{align}
\Lambda_G^{QW} (s) = \prod_{ \lambda {}_{{\bf M}_n} \in {\rm Spec} ({\bf M}_n)} \left( \lambda {}_{{\bf M}_n} - s (1-s) \right). 
\label{mankai01}
\end{align}
We recall that it follows from Eq. \eqref{wakatakakage02} that any $\rho \in {\rm Zero} (\Lambda_G^{QW})$ should be expressed as 
\begin{align}
\rho = \frac{1}{2} + \frac{i}{2} \cot \left( \frac{\theta}{2} \right).
\label{kotonowaka00}
\end{align}
If so, then $\Lambda_G^{QW} (s)$ satisfies the Riemann Hypothesis, that is, if $\rho \in {\rm Zero} (\Lambda_G^{QW})$, then $\Re (\rho) = 1/2$. Next noting Eq. \eqref{mankai01}, we need to confirm that $\rho \in {\rm Zero} (\Lambda_G^{QW})$ with Eq. \eqref{kotonowaka00} satisfies
\begin{align*}
s^2 - s + \lambda {}_{{\bf M}_n} = 0, 
\end{align*}
for $\lambda {}_{{\bf M}_n} \in {\rm Spec} ({\bf M}_n)$. This is, 
\begin{align*}
(s - \rho) (s - \overline{\rho}) =  s^2 - s + \lambda {}_{{\bf M}_n}. 
\end{align*}
So we have 
\begin{align}
\rho + \overline{\rho} &= 1
\label{kotonowaka03a}
\\
\rho \overline{\rho} &= \lambda {}_{{\bf M}_n}. 
\label{kotonowaka03b}
\end{align}
Then Eq. \eqref{kotonowaka03a} comes from Eq. \eqref{kotonowaka00}. Concerning Eq. \eqref{kotonowaka03b}, by using Eq. \eqref{kotonowaka00}, we compute 
\begin{align*}
\rho \overline{\rho}
& = \left( \frac{1}{2} + \frac{i}{2} \cot \left( \frac{\theta}{2} \right) \right) \left( \frac{1}{2} - \frac{i}{2} \cot \left( \frac{\theta}{2} \right) \right) \\
&= \frac{1}{4} \left( 1 +  \cot^2 \left( \frac{\theta}{2} \right) \right)
= \frac{1}{2} \ \frac{1}{1 - \cos \theta} 
\\
&= \frac{1}{2} \ \frac{1}{1 - \lambda {}_{{\bf P}_n}},
\end{align*}
where $\theta \in (0, 2 \pi)$. Thus we have
\begin{align}
\rho \overline{\rho} = \frac{1}{2} \ \frac{1}{1 - \lambda {}_{{\bf P}_n}}.
\label{kotonowaka04}
\end{align}
Combining Eq. \eqref{kotonowaka03b} with Eq. \eqref{kotonowaka04} yields 
\begin{align}
\lambda {}_{{\bf M}_n} = \frac{1}{2} \left( 1 - \lambda {}_{{\bf P}_n} \right)^{-1}. 
\label{kotonowaka05}
\end{align}
Then Eq. \eqref{kotonowaka05} suggests that ${\bf M}_n$ may be expressed as 
\begin{align}
{\bf M}_n =\frac{1}{2} \left( {\bf I}_n - {\bf P}_n \right)^{-1}. 
\label{wakatakakage03}
\end{align}
Therefore we define ${\rm Spec} ({\bf M}_n)$ by
\begin{align*}
{\rm Spec} ({\bf M}_n) = \left\{ \left[ \frac{1}{2} \> \frac{1}{1-\cos \theta_1} \right]^{l_1}, \ \left[ \frac{1}{2} \> \frac{1}{1-\cos \theta_2} \right]^{l_2}, \ \ldots \ , \left[ \frac{1}{2} \> \frac{1}{1-\cos \theta_p} \right]^{l_p} \right\},
\end{align*}
since 
\begin{align*}
{\rm Spec} ({\bf P}_n) = \left\{ \left[ \cos \theta_1 \right]^{l_1}, \ \left[ \cos \theta_2 \right]^{l_2}, \ \ldots \ , \left[ \cos \theta_p \right]^{l_p} \right\}.
\end{align*}
Remark that $|{\rm Spec} ({\bf M}_n) | = l_1 + l_2 + \cdots + l_p = n$ and $0 = \theta_1 < \theta_2 < \cdots < \theta_p < 2 \pi$. In particular, when $\cos \theta_1 =1$ (i.e., $\theta_1=0$), we put
\begin{align*}
\frac{1}{2} \> \frac{1}{1-\cos \theta_1} = + \infty.
\end{align*}
This matrix ``${\bf M}_n$" is exactly what we wanted to find. However, $\left( {\bf I}_n - {\bf P}_n \right)^{-1}$ does not exist, so we understand Eq. \eqref{wakatakakage03} ``in the sense of the eigenvalue", that is, as Eq. \eqref{kotonowaka05}. In fact, we see that $0 \in {\rm Spec} \left( {\bf I}_n - {\bf P}_n \right)$, since all one vector ${\bf 1}_n$ is the eigenvector corresponding to eigenvalue 0, i.e., $({\bf I}_n - {\bf P}_n) {\bf 1}_n = 0 \cdot {\bf 1}_n$. Thus ${\bf I}_n - {\bf P}_n$ is not invertible. Therefore Eq. \eqref{wakatakakage03} implies that ${\bf M}_n$ does not exist. If $G$ is $(q+1)$-regular, then we have the Laplacian ${\bf \Delta}_n$ as 
\begin{align}
{\bf \Delta}_n = {\bf D}_n - {\bf A}_n = (q+1) \left( {\bf I}_n - {\bf P}_n \right).
\label{wakatakakage04}
\end{align}
It follows from Eq. \eqref{wakatakakage04} that Eq. \eqref{wakatakakage03} becomes
\begin{align*}
{\bf M}_n = \frac{q+1}{2} \ {\bf \Delta}_n ^{-1}
\end{align*}
in the sense of the eigenvalue. We also remark that ${\bf \Delta}_n ^{-1}$ does not exist. Then we obtain the following main result.
\begin{theorem}
Let $G$ be a simple connected graph with $n$ vertices and $m$ edges. Put
\begin{align*}
\Lambda^{QW} _G (s) = \det \left( {\bf M}_n - s (1-s) {\bf I}_{n} \right),
\end{align*}
where 
\begin{align*}
{\bf M}_n =\frac{1}{2} \left( {\bf I}_n - {\bf P}_n \right)^{-1} 
\end{align*}
in the sense of the eigenvalue. Then $\Lambda^{QW} _G (s)$ satisfies 
\par
\
\par\noindent
{\rm (i)} Functional Equation:
\begin{align*}
\Lambda^{QW} _G (s) = \Lambda^{QW} _G (1- s). 
\end{align*}
\par\noindent
{\rm (ii)} Riemann Hypothesis:
\begin{align*}
\Re (\rho) = \frac{1}{2} \quad \text{for any} \ \  \rho \in {\rm Zero} (\Lambda^{QW} _G),
\end{align*}
where ${\rm Zero} (\Lambda^{QW} _G) = \left\{ s \in \CM : \Lambda^{QW} _G (s) = 0 \right\}.$
\end{theorem}

\section{Example \label{sec04}}
This section gives examples, i.e., (i) $m > n$, (ii) $m=n$, and (iii) $m < n$ cases. The results here can be obtained by the direct computation and the Konno-Sato theorem.
\par
\
\par
(i) $m > n$ case. $G=K_n$ (complete graph) with $n$ vertices and $m=n(n-1)/2$ edges for $n \ge 4$. We should remark that if $n=3$, then (ii) $m=n=3$ case. If $n=2$, then (iii) $m=1 < n=2$ case. Then we get
\begin{align*}
{\rm Spec} \left( {\bf P}_n \right) 
&= \left\{ [1]^{1}, \  \left[ - \frac{1}{n-1} \right]^{n-1} \right\}, 
\\
{\rm Spec} \left( {\bf U}_{n(n-1)} \right)
&= \Biggl\{ [1]^{(n(n-3)+4)/2}, \  [-1]^{n(n-3)/2}, \ 
\\
& \qquad \qquad \qquad \left[ \frac{-1 + i \sqrt{n(n-2)}}{n-1} \right]^{n-1}, \ \left[ \frac{-1 - i \sqrt{n(n-2)}}{n-1} \right]^{n-1} \Biggr\}. 
\end{align*}
Remark that we confirm that spectral mapping theorem holds such as 
\begin{align*}
\Re \left[ \frac{-1 + i \sqrt{n(n-2)}}{n-1} \right] = \Re \left[ \frac{-1 - i \sqrt{n(n-2)}}{n-1} \right] = - \frac{1}{n-1}.
\end{align*}
Therefore we have
\begin{align*}
{\rm Zero} \left( \Lambda_{K_n}^{QW} \right) 
&= \left\{ \ \left[ \ \frac{1}{2} + i \cdot (+ \infty) \right]^2, \ \left[ \ \frac{1}{2} + \frac{i}{2} \sqrt{1 - \frac{2}{n}} \  \right]^{n-1}, \ \left[ \ \frac{1}{2} - \frac{i}{2} \sqrt{1 - \frac{2}{n}} \  \right]^{n-1} \ \right\}
\\
&\cup \left\{ \ \left[ \frac{1}{2} + i \cdot ( + \infty) \right]^{n(n-3)/2}, \ \left[ \frac{1}{2} \right]^{n(n-3)/2} \right\}.
\end{align*}
Note that $m-n = n(n-3)/2$. Furthermore, we obtain
\begin{align*}
{\rm Spec} \left( {\bf M}_n \right) 
= \left\{ \left[ \ + \infty \  \right]^1,  \left[ \ \frac{1}{2} \left( 1 - \frac{1}{n} \right) \  \right]^{n-1} \right\},
\end{align*}
for $n \ge 4$. Formally, if we take $n \to \infty$, then 
\begin{align*}
\lim_{n \to \infty} {\rm Spec} \left( {\bf M_{n}} \right) 
= \left\{ \left[ \ + \infty \  \right]^1,  \left[ \ \frac{1}{2} \  \right]^{+ \infty} \right\}.
\end{align*}
\par
\
\par
(ii) $m = n$ case. $G=C_n$ (cycle graph) with $n$ vertices and $m=n$ edges. In this case, $2n \times 2n$ matrix ${\bf U}_{2n}$ and $n \times n$ matrix ${\bf P}_{n}$ are expressed as follows (see \cite{K1, K2}, for example):
\begin{align*}
{\bf U}_{2n} =
\begin{bmatrix}
O & P & O & \dots & \dots & O & Q \\
Q & O & P & O & \dots & \dots & O \\
O & Q & O & P & O & \dots & O \\
\vdots & \ddots & \ddots & \ddots & \ddots & \ddots & \vdots \\
O & \dots & O & Q & O & P & O \\
O & \dots & \dots & O & Q & O & P \\
P & O & \dots & \dots & O & Q & O
\end{bmatrix} 
, \quad
{\bf P}_n = \frac{1}{2}
\begin{bmatrix}
0 & 1 & 0 & \dots & \dots & 0 & 1 \\
1 & 0 & 1 & 0 & \dots & \dots & 0 \\
0 & 1 & 0 & 1 & 0 & \dots & 0 \\
\vdots & \ddots & \ddots & \ddots & \ddots & \ddots & \vdots \\
0 & \dots & 0 & 1 & 0 & 1 & 0 \\
0 & \dots & \dots & 0 & 1 & 0 & 1 \\
1 & 0 & \dots & \dots & 0 & 1 & 0
\end{bmatrix} 
,
\end{align*}
where
\begin{align*}
P=
\begin{bmatrix}
1 & 0 \\
0 & 0 
\end{bmatrix} 
,
\quad
Q=
\begin{bmatrix}
0 & 0 \\
0 & 1 
\end{bmatrix} 
,
\quad
O=
\begin{bmatrix}
0 & 0 \\
0 & 0 
\end{bmatrix} 
.
\end{align*}
Let $\xi_k = 2 \pi k/n$ for $k=0,1, \ldots , n-1$. Thus we get
\begin{align*}
{\rm Spec} \left( {\bf P}_n \right) 
&= \left\{ [\cos \xi_k ]^{1} \ : \ k=0,1, \ldots , n-1 \right\}, 
\\
{\rm Spec} \left( {\bf U}_{2n} \right)
&= \left\{ [e^{i \xi_k} ]^{1}, \  [e^{- i \xi_k} ]^{1} \ : \ k=0,1, \ldots , n-1 \right\}. 
\end{align*}
Therefore we obtain
\begin{align*}
{\rm Zero} \left( \Lambda_{C_n}^{QW} \right)
= \left\{ \ \left[ \ \frac{1}{2} + \frac{i}{2} \cot \left( \frac{\xi_k}{2} \right)  \  \right]^1 , \ \left[ \ \frac{1}{2} - \frac{i}{2} \cot \left( \frac{\xi_k}{2} \right)  \  \right]^1 \ : \ k=0,1, \ldots , n-1 \right\}.
\end{align*}
Formally, if we take $n \to \infty$, then 
\begin{align*}
\lim_{n \to \infty} {\rm Zero} \left( \Lambda_{C_n}^{QW} \right)
= \left\{ \ \left[ \ \frac{1}{2} + i \cdot \gamma \  \right]^1 \ : \ \gamma \in \RM \ \right\}.
\end{align*}
Moreover we get
\begin{align*}
{\rm Spec} \left( {\bf M}_n \right) 
= \left\{ \ \left[ \ + \infty \  \right]^1, \ \left[ \frac{1}{2} \> \frac{1}{1-\cos \xi_1} \right]^1, \ \ldots, \ \left[ \frac{1}{2} \> \frac{1}{1-\cos \xi_{n-1}} \right]^1 \ \right\},
\end{align*}
since $\xi_0 =0$ (i.e., $\cos \xi_0 =1$). From now on, we consider $n=3$ and $n=4$ cases. First we deal with $n=3$. Then we have
\begin{align*}
{\rm Spec} \left( {\bf P}_3 \right) 
&= \left\{ \left[ \cos \left( 0 \cdot \pi/3 \right) \right]^{1}, \ \left[ \cos \left( 2 \cdot \pi/3 \right) \right]^{1}, \ \left[ \cos \left( 4 \cdot \pi/3 \right) \right]^{1} \right\} 
= \left\{ [1]^1, \ \left[ - \frac{1}{2} \right]^2 \right\}, 
\\
{\rm Spec} \left( {\bf U}_6 \right)
&= \left\{ \left[ e^{i (0 \cdot \pi/3) } \right]^{2}, \ \left[ e^{i (2 \cdot \pi/3) } \right]^{2}, \ \left[ e^{i (4 \cdot \pi/3) } \right]^{2} \right\}
\\
&
= \left\{ [1]^2, \ \left[\frac{-1+i \sqrt{3}}{2} \right]^2, \ \left[\frac{-1-i \sqrt{3}}{2} \right]^2 \right\}.
\end{align*}
Thus we obtain
\begin{align*}
{\rm Zero} \left( \Lambda_{C_3}^{QW} \right) 
= \left\{ \left[\frac{1}{2} + i \cdot ( +\infty) \right]^2, \ \left[\frac{1}{2} + \frac{i \sqrt{3}}{6} \right]^2, \ \left[\frac{1}{2} - \frac{i \sqrt{3}}{6} \right]^2 \right\}.
\end{align*}
Note that 
\begin{align*}
{\rm Spec} \left( {\bf M}_3 \right) 
= \left\{ \left[ + \infty \right]^{1}, \ \left[ \frac{1}{3} \right]^{2} \right\}, \qquad {\rm Spec} \left( {\bf \Delta}_3 \right) = \left\{ \left[ 0 \right]^{1}, \ \left[ 3 \right]^{2} \right\},
\end{align*}
where
\begin{align*}
{\bf \Delta}_3 
= 2 \left( {\bf I}_3 - {\bf P}_3 \right)
=
\begin{bmatrix}
2 & -1 & -1 \\
-1 & 2 & -1 \\
-1 & -1 & 2 
\end{bmatrix}
.
\end{align*}
So we confirm that ${\bf M}_3 = {\bf \Delta}_3 ^{-1}$ does not exist. As in the case of $n=3$, we treat $n=4$ case. Then we get
\begin{align*}
{\rm Spec} \left( {\bf P}_4 \right) 
&= \left\{ \left[ \cos \left( 0 \cdot \pi/4 \right) \right]^{1}, \ \left[ \cos \left( 2 \cdot \pi/4 \right) \right]^{1}, \ \left[ \cos \left( 4 \cdot \pi/4 \right) \right]^{1} , \ \left[ \cos \left( 6 \cdot \pi/4 \right) \right]^{1} \right\} \\
&
= \left\{ [1]^1, \ [0]^2, \ [-1]^1 \right\}, 
\\
{\rm Spec} \left( {\bf U}_8 \right)
&= \left\{ \left[ e^{i (0 \cdot \pi/4) } \right]^{2}, \ \left[ e^{i (2 \cdot \pi/4) } \right]^{2}, \ \left[ e^{i (4 \cdot \pi/4) } \right]^{2}, \ \left[ e^{i (6 \cdot \pi/4) } \right]^{2} \right\}
\\
&
= \left\{ [1]^2, \ \left[ i \right]^2, \ \left[ -1 \right]^2, \ \left[ -i \right]^2 \right\}.
\end{align*}
Therefore we have
\begin{align*}
{\rm Zero} \left( \Lambda_{C_4}^{QW} \right)  
= \left\{ \left[\frac{1}{2} + i \cdot ( +\infty) \right]^2, \ \left[\frac{1}{2} + \frac{i}{2} \right]^2, \  \left[\frac{1}{2} \right]^2, \ \left[\frac{1}{2} - \frac{i}{2} \right]^2 \right\}.
\end{align*}
Remark that
\begin{align*}
{\rm Spec} \left( {\bf M}_4 \right) 
= \left\{ \left[ + \infty \right]^{1}, \ \left[ \frac{1}{2} \right]^{2}, \ \left[ \frac{1}{4} \right]^{1} \right\}, \qquad {\rm Spec} \left( {\bf \Delta}_4 \right) = \left\{ \left[ 0 \right]^{1}, \ \left[ 2 \right]^{2}, \ \left[ 4 \right]^{1} \right\},
\end{align*}
where
\begin{align*}
{\bf \Delta}_4 = 2 \left( {\bf I}_4 - {\bf P}_4 \right)
=
\begin{bmatrix}
2 & -1 & 0 & -1 \\
-1 & 2 & -1 & 0 \\
0 & -1 & 2 & -1 \\
-1 & 0 & -1 & 2 
\end{bmatrix}
.
\end{align*}
Thus we confirm that ${\bf M}_4 = {\bf \Delta}_4 ^{-1}$ does not exist. 
\par
\
\par
(iii) $m < n$ case. $G=S_n$ (star graph) with $n$ vertices and $m=n-1$ edges. Remark that $S_n$ is isomorphic to the complete bipartite graph $K_{1,n-1}$. Then we obtain
\begin{align*}
{\rm Spec} \left( {\bf P}_n \right) 
&= \left\{ [1]^1, \ [0]^{n-2}, \ [-1]^1 \right\}, 
\\
{\rm Spec} \left( {\bf U}_{2(n-1)} \right)
&= \left\{ [1]^2, \ \left[ i \right]^{n-2}, \ \left[ -i \right]^{n-2}, \ \left[ -1 \right]^2 \right\} \setminus \left\{ [1]^1, \ \left[ -1 \right]^1 \right\}.
\\
&=\left\{ [1]^1, \ \left[ i \right]^{n-2}, \ \left[ -i \right]^{n-2}, \ \left[ -1 \right]^1 \right\}. 
\end{align*}
Thus we have
\begin{align*}
{\rm Zero} \left( \Lambda_{S_n}^{QW} \right)  
= \left\{ \left[\frac{1}{2} + i \cdot ( +\infty) \right]^1, \ \left[\frac{1}{2} + \frac{i}{2} \right]^{n-2}, \  \left[\frac{1}{2} - \frac{i}{2} \right]^{n-2}, \ \left[\frac{1}{2} \right]^1 \right\}.
\end{align*}
Note that
\begin{align*}
{\rm Spec} \left( {\bf M}_n \right) 
= \left\{ \left[ + \infty \right]^{1}, \ \left[ \frac{1}{2} \right]^{n-2}, \ \left[ \frac{1}{4} \right]^{1} \right\}, \qquad {\rm Spec} \left( {\bf \Delta}_n \right) = \left\{ \left[ 0 \right]^{1}, \ \left[ 2 \right]^{n-2}, \ \left[ 4 \right]^{1} \right\}.
\end{align*}

\section{Conclusion \label{sec05}}
In this paper, we introduced a new zeta function $\Lambda^{QW} _G (s) = \det \left( {\bf M}_n - s (1-s) {\bf I}_{n} \right)$ for a suitable $n \times n$ matrix ${\bf M}_n$ on a simple connected graph $G$ with $n$ vertices via the QW by the help of the Konno-Sato theorem. In addition, we showed that $\Lambda^{QW} _G (s)$ satisfies the Functional Equation: $\Lambda^{QW} _G (s) = \Lambda^{QW} _G (1- s)$ and the Riemann Hypothesis: if $\rho \in {\rm Zero} (\Lambda^{QW} _G)$, then $\Re (\rho) = 1/2$. The challenging problem, of course, is to prove the original Riemann Hypothesis: if $\rho \in {\rm Zero} (\Lambda)$, then $\Re (\rho) = 1/2$, by using our approach based on the QW. Moreover, one of the interesting problems might be to clarify the relation between our zeta function $\Lambda_G^{QW} (s)$ and other zeta functions for a graph $G$.





\begin{thebibliography}{99}

\bibitem{Andrews1999} 
Andrews, G. E., Askey, R., Roy, R.: 
Special Functions. Cambridge University Press (1999)


\bibitem{Bass}
Bass, H.: 
The Ihara-Selberg zeta function of a tree lattice. 
Internat. J. Math. {\bf 3}, 717-797 (1992)


\bibitem{Edwards}
Edwards, H. M.:
Riemann's Zeta Function.
Academic Press (1974)




\bibitem{GM}  
Gelbart, S. S., Miller, S. D.:
Riemann's zeta function and beyond.
Bull. Amer. Math. Soc. {\bf 41}, 59--112 (2003)


\bibitem{Ihara}
Ihara, Y.: 
On discrete subgroups of the two by two projective linear group 
over $p$-adic fields. 
J. Math. Soc. Japan {\bf 18}, 219--235 (1966)


\bibitem{K0}
Komatsu, T., Konno, N., Sato, I.: 
A note on the Grover walk and the generalized Ihara zeta function of the one-dimensional integer lattice. 
Yokohama Math. J. 
{\bf 67}, 115--123 (2021).

\bibitem{K1}
Komatsu, T., Konno, N., Sato, I.: 
Grover/Zeta Correspondence based on the Konno-Sato theorem.
Quantum Inf. Process. {\bf 20}, 268 (2021) 


\bibitem{K2}
Komatsu, T., Konno, N., Sato, I.: 
Walk/Zeta Correspondence.
arXiv:2104.10287 (2021) 


\bibitem{K3}
Komatsu, T., Konno, N., Sato, I.: 
IPS/Zeta Correspondence.
Quantum Inf. Comput. 
{\bf 22}, 251--269 (2022)


\bibitem{K4}
Komatsu, T., Konno, N., Sato, I.: 
Vertex-Face/Zeta Correspondence.
J. Algebraic Combin. (in press)
arXiv:2107.03300 (2021) 


\bibitem{K5}
Komatsu, T., Konno, N., Sato, I.: 
CTM/Zeta Correspondence. 
Quantum Stud.: Math. Found. 
{\bf 9}, 165--173 (2022)


\bibitem{K7}
Komatsu, T., Konno, N., Sato, I., Tamura, S.: 
A Generalized Grover/Zeta Correspondence. 
arXiv:2201.03973 (2022) 


\bibitem{K8}
Komatsu, T., Konno, N., Sato, I., Tamura, S.: 
Mahler/Zeta Correspondence. 
arXiv:2202.05966 (2022)


\bibitem{Konno2008}
Konno, N.: 
Quantum Walks. In: Quantum Potential Theory, Franz, U., and Schurmann,
M., Eds., Lecture Notes in Mathematics: Vol. 1954, pp.309--452, Springer-Verlag, Heidelberg (2008)


\bibitem{KonnoSato} 
Konno, N., Sato, I.: 
On the relation between quantum walks and zeta functions. 
Quantum Inf. Process. {\bf 11}, 341--349 (2012)


\bibitem{K6}
Konno, N., Tamura, S.: 
Walk/Zeta Correspondence for quantum and correlated random walks. 
Yokohama Math. J. {\bf 67}, 125--152 (2021)


\bibitem{ManouchehriWang}
Manouchehri, K., Wang, J.: 
Physical Implementation of Quantum Walks.
Springer, New York (2014)


\bibitem{Norris}
Norris, J. R.: Markov Chains. 
Cambridge University Press, Cambridge (1997)


\bibitem{Portugal} 
Portugal, R.: 
Quantum Walks and Search Algorithms, 2nd edition. 
Springer, New York (2018)


\bibitem{RenEtAl}  
Ren, P., Aleksic, T., Emms, D., Wilson, R. C., Hancock, E. R.: 
Quantum walks, Ihara zeta functions and cospectrality in regular graphs.  
Quantum Inf. Process. {\bf 10}, 405--417 (2011)   


\bibitem{SS}  
Segawa, E., Suzuki, A.: 
Spectral mapping theorem of an abstract quantum walk.  
Quantum Inf. Process. {\bf 18}, 333 (2019)   


\bibitem{Spitzer}  
Spitzer, F.: 
Principles of Random Walk, 2nd edition. 
Springer, New York (1976)


\bibitem{Terras}  
Terras, A.: 
Zeta Functions of Graphs. 
Cambridge Univ. Press (2011)


\bibitem{Venegas} 
Venegas-Andraca, S. E.: 
Quantum walks: a comprehensive review. 
Quantum Inf. Process. {\bf 11}, 1015--1106 (2012)




\end{thebibliography}
\end{document}